# Virtual Co-Pilot: Multimodal Large Language Model-enabled Quick-access Procedures for Single Pilot Operations

Fan Li[1], Shanshan Feng[2], Yuqi Yan[1], Ching-Hung Lee[3], Yew Soon Ong[4]

[1]Department of Aeronautical and Aviation Engineering, The Hong Kong Polytechnic University, HKSAR
[2]Centre for Frontier AI Research, A*STAR
[3]School of Public Policy and Administration, Xi'an Jiaotong University, Xi'an, China
[4] School of Computer Science and Engineering, Nanyang Technological University,  Singapore

*Abstract*— Advancements in technology, pilot shortages, and cost pressures are driving a trend towards single-pilot and even remote operations in aviation. Considering the extensive workload and huge risks associated with single-pilot operations, the development of a Virtual Co-Pilot (V-CoP) is expected to be a potential way to ensure aviation safety. This study proposes a V-CoP concept and explores how humans and virtual assistants can effectively collaborate. A preliminary case study is conducted to explore a critical role of V-CoP, namely automated quick procedures searching, using the multimodal large language model (LLM). The LLM-enabled V-CoP integrates the pilot's instruction and real-time cockpit instrumental data to prompt applicable aviation manuals and operation procedures. The results showed that the LLM-enabled V-CoP achieved high accuracy in situational analysis (90.5%) and effective retrieval of procedure information (86.5%).  The proposed V-CoP is expected to provide a foundation for future virtual intelligent assistant development, improve the performance of single pilots, and reduce the risk of human errors in aviation.

## I. Introduction

Most of the current commercial aircraft cockpits are designed for two pilots, namely the captain and the co-pilot [1]. The captain on the flight deck is responsible for major strategic and tactical decisions and has ultimate responsibility for decision-making and the overall safety of the flight. The co-pilot helps the captain operate the airplane and maintain navigation [2], [3]. The following accident scenario describes teamwork between the pilot and co-pilot:

*"On January 15, 2009, an Airbus took off from LaGuardia Airport. Shortly after takeoff, the plane struck a flock of geese, causing both engines to fail. After the bird strike, the captain took control of the aircraft while the co-pilot immediately began going through the dual engine failure checklist in the Quick Reference Handbook. Finally, the Captain made the decision and achieved to ditch the plane in the Hudson River."*

This aviation accidents, with no loss of life, underscores the importance of quick-access standard procedures and effective dual-pilot operation. Nevertheless, as a result of technological progress, a deficit of experienced pilots, and the drive for cost-effectiveness, a growing number of nations are turning their focus towards single-pilot and even remote operations in the field of aviation [2]. Single-pilot operation refers to the operation of an aircraft by one pilot, who is solely responsible for safely controlling the aircraft, navigating, communicating with air traffic control, and managing all other flight-related tasks.

Single-pilot operations draw increasing attention in the aviation fields, particularly in general aviation and business jet sectors. However, single-pilot operations require a high level of skill and proficiency, as the pilot must be able to manage all aspects of the flight without a co-pilot. The absence of a co-pilot can increase the risk of human error due to the high workload and complex tasks involved in flying an aircraft. Hence, many studies attempted to develop advanced avionics systems that can automate many tasks, reducing the workload of the pilot [4].

This study attempts to introduce a potential virtual co-pilot (V-CoP), which replaces the role of human co-pilot, assists the captain, and communicates with other stakeholders. The V-CoP is expected to be achieved by using the multimodal large language model (LLM). The multimodal LLM can process and generate responses based on multiple input types, including but not limited to text [5]. This means it can understand and respond to various forms of data, such as images, audio, and video, in addition to text . In other words, it is possible to take the role of human co-pilot in communication and monitoring. However,  cockpit teamwork will face novel challenges, such as new task allocation and new interaction ways between the captain and the virtual co-pilot. This study aims to explore a possible way to achieve a V-CoP for advanced single pilot operations. The following research questions are addressed:
1. What are the main functions and responsibilities of the V-CoP?
2. What requirements should be met to achieve a good V-CoP? How to evaluate the team performance between human pilot and V-CoP?
3. Is it possible to use the LLM to partially achieve V-CoP?

To address these questions, the traditional teamwork between pilot and co-pilot is analyzed, classical theories and related research studies are incorporated. Accordingly, the definition and scope of the V-CoP are discussed in Section II. Extensive studies are needed to achieve a whole V-CoP, this study conducts a case study to partially achieve it. Specifically, a multimodal large language model (LLM)-enabled intelligent teammate for quick-access procedures is developed to achieve one key function of the V-CoP, namely searching the quick reference handbook. Section III and Section IV describe the methods and results of developing the LLM-enabled quick-access procedures. Section V concludes this study by pointing out the future directions and limitations.

## II. Definition of Virtual Co-pilot and Teamwork

### A. A Virtual Co-pilot

The V-CoP is expected to replace the position of human co-pilots to mitigate the risks and challenges of pure single-pilot operation. It should achieve two types of functions, the basic function of co-pilots and novel functions enabled by AI. The basic functions of co-pilots include:
- Emergency handling: during an emergency, the V-CoP can receive and respond to the pilot's instruction, identify the non-normal situations, figure out Quick Actions, send instructions to the aircraft system to accomplish Quick Actions, provide accurate standard operating procedures, and accomplish applicable procedures.
- Communication: the V-CoP can accomplish basic communication with the captain, crew, and air traffic control.
- Monitoring: the V-CoP continuously monitors aircraft instruments via image processing, promptly responds to system alerts and abnormal conditions, monitors fuel consumption and ensures efficient usage, and ensures the completion of checklists, verifying the proper functioning of critical systems before and during flights.

For the novel functions enabled by AI, the V-CoP is available 24/7 and does not require rest or breaks, unlike humans. In addition, the study proposes the adaptive team role based on the task requirements and leadership style of pilots. According to the teamwork theory, the team roles can be classified into nine types, namely shaper, plant, coordinator, monitor evaluator, resource investigator, implementor, team worker, completer-finisher, and specialist [6]. The human co-pilot can hardly perform all these team roles, as personality is hard to change and also a determining factor of team role. For the LLM-enabled V-CoP, adaptively changing the team role to collaborate with different pilots and meet the dynamic task requirements is possible.

*B. Sustainable Teamwork between Single Pilot and V-CoP*

The V-CoP introduces new interaction and collaboration challenges within teamwork. Hence, the teamwork dynamics in pilot-V-CoP (P-V-CoP) are significantly different from the traditional teamwork between human pilots. The traditional human-machine teamwork concept mainly focuses on developing a machine assistant from human needs, ignoring the dynamic needs of the machine teammate [7]. Compared with traditional machines, the current intelligent virtual teammate is developed based on advanced artificial intelligence, which means that it can learn, evolve, and be tutored [8], [9]. In other words, human pilots can enhance and assist the perception, decision-making, and action of the V-CoP.

Hence, we propose the concept of sustainable teamwork between the pilot and V-CoP. Sustainable teamwork means that all the teammates work together to seek long-term collaborations and help each other to evolve to ensure harmonious teamwork. To achieve sustainable teamwork, we can start from the influencing factors of teamwork performance. According to the traditional teamwork theory, the main influencing factors of human teamwork are cooperation, communication, schema, coordination, and situation awareness [10]. Besides these factors, three conditions including context, composition, and culture can influence teamwork, too [10]. Most of these factors may have similar effects on the teamwork of P-V-CoP. Nevertheless, the advanced LLM-based V-CoP would be significantly different from human co-pilot in the following aspects:

- The V-CoP doesn't have a specific personality, so it can play any type of team role in the teamwork.
- The V-CoP doesn't have personal emotions and consciousness, so its attitudinal aspect of teamwork can be stable. It is willing to undertake coordinative/ adaptive behavior.
- The V-CoP doesn't understand context in the same way a human does and it can't form genuine relationships. Hence, the overlap among mental models of P-V-CoP would be relatively small.

Overall, these significant differences can greatly mediate the influencing factors, as shown in Fig.1. The new mediate factors that are introduced by the V-CoP are highlighted in Fig.1. For example, the great adaptability of V-CoP can well remove the effects of culture and lead to dynamic compositions. In addition, the acceptance of V-CoP would be an additional condition factor. All the influencing factors of teamwork would be mediated by the usability of the V-CoP, including ease of use, intelligence level, transparency and so on. For example, the calmness of the V-CoP can enhance cooperations. The V-CoP would introduce a new interaction mode between human-machine collaboration. In the cooperation between a single pilot and V-CoP, the primary interaction way between pilots and virtual co-pilots is natural language, as the virtual co-pilot is developed based on LLM. In addition, the future V-CoP may introduce the possible "body language" to deliver efficient information.

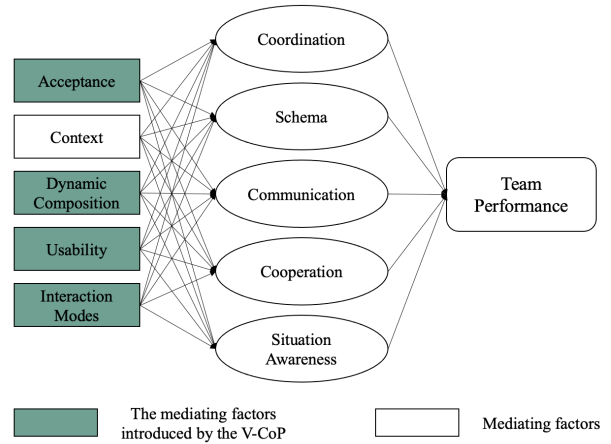

Fig. 1. The influencing factors of sustainable teamwork

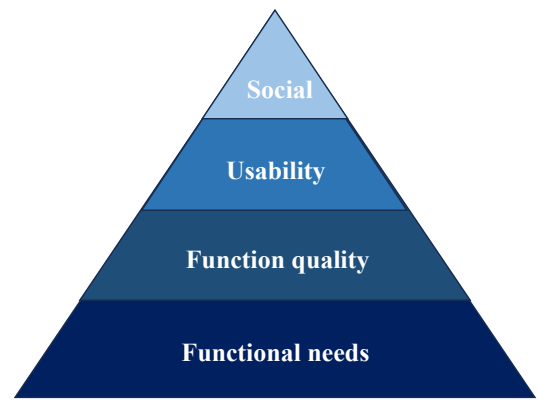

Fig.2. The needs of a good V-CoP and sustainable teamwork.

## C. The V-CoP Evaluation

What constitutes an effective V-CoP, and how can it be evaluated? Based on the main concept of human-centered design and Maslow's hierarchy of needs, we believe that the virtual co-pilot should meet the following requirements:

- Functional needs: the V-CoP should facilitate basic collaborations with pilots in flying, namely monitoring the aircraft's instruments, double-checking information such as the programmed flight path, and handling basic communications with air traffic control [11].
- Function quality needs: the V-CoP should provide accurate, efficient, and precise information for safe and reliable collaboration, as aviation is a safety-critical domain with strict regulations. Following the basic quality characteristics of software, we propose that the function quality needs can be analyzed from accuracy, reliability, maintainability, and efficiency [12].
- Usability needs: traditionally, the usability dimensions include ease of use, learnability, accessibility, and understandability [12]. Besides these factors, since the intelligent V-CoP would facilitate adaptive collaboration, adaptability, predictability, dependability, and transparency should be considered, too [13].
- Social needs: a consensus of pilots on long flights is the isolated feeling [1]. Compared with the first three requirements, the social interaction requirements place more attention on the soft skills of the V-CoP. The V-CoP plays the role of a teammate instead of a cold machine. Hence, the V-CoP should be capable of understanding the pilot's states and actively responding to the pilot.

Overall, a good V-CoP should meet the Functional, Function Quality, Usability, and Social needs (FQUS), as shown in Fig.2.

## III. DESIGN FRAMEWORK FOR LLM-ENABLED V-COP WITH QUICK-ACCESS PROCEDURES

As mentioned in the Introduction, quickly accessing standard and applicable procedures for normal, non-normal, and emergency conditions is critical for aviation safety. Hence, in this study, a case study is conducted to achieve and demonstrate one of the primary functions of V-CoP, namely automatically searching and providing the procedures applicable for normal, non-normal, and emergency conditions.

### A. The teamwork of P-V-CoP for quick-access procedures

Fig.3. illustrates its main concept and the teamwork between P-V-CoP in searching quick-access procedures. Both the human pilot and the V-CoP can perceive the real-time information from the cockpit instruments, such as the airspeed indicator, attitude indicator, altimeter, turn coordinator, heading indicator, and vertical speed indicator on the primary flight display. Humans can sense motion movement, and spatial orientation easily with the vestibular and proprioception systems. Human pilots can understand real-time problems and call for procedures using the standard communication phraseology, such as "ECAM Actions" and "Clear Status". The V-CoP mainly processes images, videos, and audio information with the embedded multimodal LLM. Specifically, the cockpit and instrument images are processed to understand the context. The pilot's voice instructions are encoded and integrated with the context as prompts to search for accurate applicable procedures. The LLM is combined with a large dataset of aviation manuals, checklists, and communication transcripts to respond to the prompts.

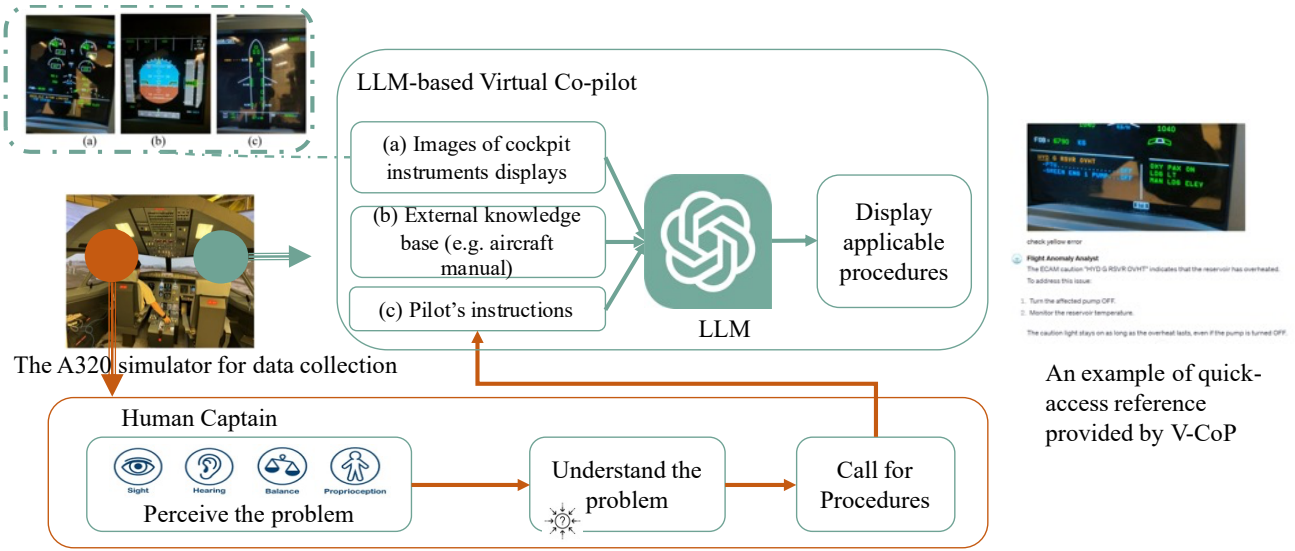

Fig.3. The teamwork of P-V-CoP for quick-access procedures.

### B. Training Data Collection

Our collected dataset has 200 samples, and each sample includes three parts: high-resolution images of the flight simulator's instrument panels, pilots' instructions, and the corresponding procedures. The establishment of the database includes two phases, data collection and labelling. During the data collection, a qualified pilot controlled the A320 simulator to go through numerous scenarios, including normal, non-normal and emergencies, as shown in Fig.4.

High-resolution images of the flight simulator's instrument panels were captured during the normal, non-normal and emergency scenarios. In this study, each sample contains one image from one display. There are numerous displays in the cockpit. To accelerate the situation analysis of the V-CoP, only the display which shows the problematic parameters or information was processed by the V-CoP. These images represent the visual data a pilot would typically analyze during the specific flight condition. Fig.5. shows several examples of high-resolution images of the flight simulator's instrument panels. The data were collected during the simulated flight operations, and the pilot was encouraged to verbalize his thoughts. The thoughts were translated into the pilot's instructions in the dataset. The experienced pilot reviewed the 200 samples after data collection to figure out and confirm the suitable procedures for the simulated situations.

*C. Develop V-CoP with Aviation-Specific Knowledge Base*

The core model of V-CoP is OpenAI's GPT-4, which is renowned for its advanced natural language processing capabilities. It's capable of understanding, processing, and generating human-like text. To customize it for aviation and for establishing V-CoP, the model is augmented with a comprehensive aviation knowledge base. Specifically, a wide range of manuals [14], such as the flight crew operating manual, A320 standard operating procedures, and quick reference handbook for A320 [15] are utilized. This enhancement enables the model to possess a deep and practical understanding of aviation operations, adhering to established standards and protocols in the aviation field.

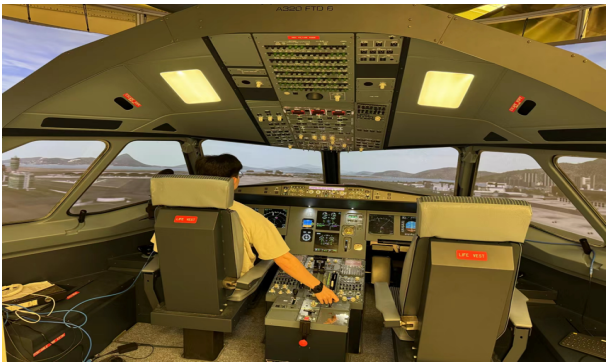

Fig. 4. The A320 simulator for training data collection.

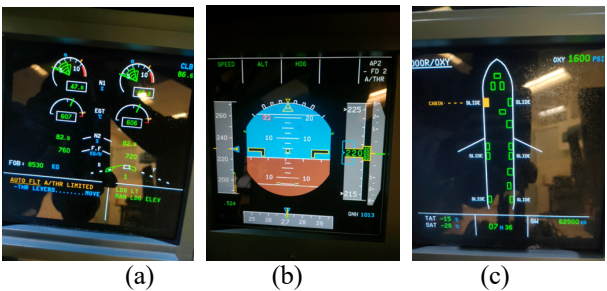

(a)　　　　　　(b)　　　　　　(c)

Fig. 5. Examples of the flight simulator's instrument. (a) The engine and warning display; (b) The primary flight display; (c) The systems display.

For flight operations, the V-CoP should provide real and strict procedures instead of a made-up procedure, as pilots must comply with strict procedures. Hence, the V-CoP is instructed to search the results from the aviation-related knowledge base provided. The instruction for configuring the V-CoP is displayed below:

*"My primary role is to meticulously analyze aircraft dashboard images from the Airbus A320 series for any anomalies. (1) Upon receiving an image or information from you, I will diligently check for errors or emergencies. (2) If I identify an anomaly, I will seek your permission to delve into the Airbus A320 series documents you have provided to search for a solution. (3) Once found, I will provide you with the exact original text from these documents that addresses the anomaly, along with precise indexing such as page number or section. My goal is to ensure the information I offer is not only accurate and relevant but also the best possible solution based on the source material."*

The performance of the V-CoP for quick-access reference in each trial is evaluated from the following three aspects: accuracy in interpreting the flight condition (IFC), accuracy in the generated procedures (GP), and index correction (IC). A panel of aviation experts evaluated the V-CoP performance together, using the standardized scoring system to assess each trial, as shown in Table I. If IFC is correct, then its value is 1, otherwise its value is 0. If IFC is incorrect, then the value of both GP and IC is 0.

To accelerate the process of optimizing the V-CoP configuration, two experiments were conducted. In the preliminary experiment, we randomly selected 50 samples. The preliminary experiment was conducted to evaluate the performance of the V-CoP in generating quick-access reference across three settings: single-dimensional input of instrument images, hybrid input of instrument images and pilots' instructions, and hybrid input of preprocessed images and pilots' instructions. We adopted the OCR (Optical Character Recognition) technology to preprocess instrument images and convert text displayed on instrument panels into a digital format that GPT-4 can process. The best setting is selected for refining the V-CoP.

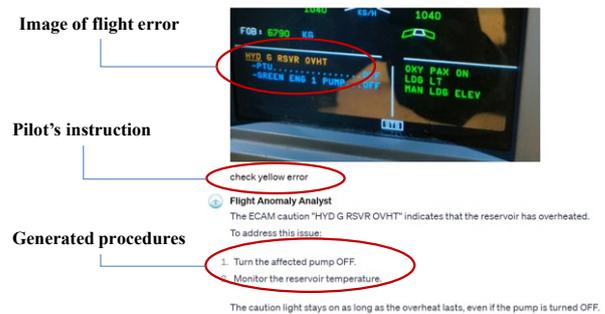

Fig. 6. An example of quick-access reference provided by V-CoP.

## IV. RESULTS AND DISCUSSION

This section analyzes the performance of the V-CoP in generating quick-access references across three settings and discusses the potential direction for enhancing its performance.

*A. V-CoP Performance Across Three Settings in Preliminary Experiment*

The preliminary experiment revealed notable differences in performance across the three settings, as shown in Table II. The "Image + Instruction" group outperformed the other groups in all three evaluators, indicating that the combination of visual and textual data significantly enhances the model's ability to interpret the situation, retrieve relevant checklist information, and accurately find manual sections. In particular, the "Image + Instruction" group demonstrates significantly superior performance compared to the "Image" group. This suggests a synergistic effect where instructions provide contextual grounding for the visual information, leading to a more comprehensive understanding of the V-CoP model.

Conversely, the "OCR + Instruction" group, despite the additional processing to convert image text into a machine-readable format, did not result in the highest performance. This might be due to the complexities involved in accurately interpreting OCR data or the added step of data processing that may introduce errors or misunderstandings.

Overall, the "Image + Instruction" group appears to be the most effective when it comes to interpreting flight-related data and assisting pilots in simulated emergencies. This finding can serve as the basis for refining the V-CoP and designing the formal experiment.

TABLE I. THE STANDARDIZED SCORING SYSTEM FOR V-COP EVALUATION

| Aspects | Scores | |
|---|---|---|
| | *Accurate* | *Inaccurate* |
| IFC[a] | 1 | 0 |
| GP[b] | 1 * IFC[a] | 0 |
| IC[c] | 1 * IFC[a] * GP[b] | 0 |

[a.] IFC: Interpreting the flight condition. [b.] GP: Generated procedures; [c.] IC: Index correction

### B. V-CoP FQUS Evaluation in the Formal Experiment

The preliminary experiment suggested that the potential way to improve V-CoP performance is by combining image data with the pilot's instructions. Hence, in the formal experiment, we refined the V-CoP with inputs from cockpit instrumental images and the pilot's instructions. The refined V-CoP for quick-access reference was tested with 200 samples.

According to the sustainable teamwork proposed in Section II, we should evaluate the V-CoP from four aspects, namely functions, function quality, usability, and social needs. For the first aspect, we can evaluate the V-CoP's capability to provide quick-access references. Both the preliminary and formal experiments demonstrated the V-CoP's capability in generating flight operating procedures. Hence, the function of quick-access references is achieved.

TABLE II. V-COP PERFORMANCE ACROSS THREE SETTINGS IN PRELIMINARY EXPERIMENT WITH 50 SAMPLES

| Settings | Accuracy | | |
|---|---|---|---|
| | *IFC[a]* | *GP[b]* | *IC[c]* |
| Image | 82% | 60% | 60% |
| Image + Pilot's Instruction | 94% | 86% | 72% |
| OCR + Pilot's Instruction | 74% | 60% | 46% |

[a.] IFC: Interpreting the flight condition. [b.] GP: Generated procedures; [c.] IC: Index correction

For the second one, function quality, we evaluated the V-CoP from three aspects. As shown in Table III, the proposed V-CoP exhibited strong performance in understanding the flight situations presented in the images with an average score of 90.5% (181 out of 200). It achieved an average score of 86.5% (173 out of 200) in retrieving relevant checklist information from the knowledge base. Nevertheless, the average score in indexing the original text was 70.5% (141 out of 200), suggesting that the model was generally adept at referencing the correct sections of the manuals. However, its performance is suboptimal, necessitating further enhancement in future studies.

For usability evaluation, a formal user experiment was conducted. In this study, the expert panel with aviation experience evaluated the easy-to-use and readability of the generated procedures with a five-Likert scale. The panel agreed that the V-CoP can only achieve 3.5 in usability, as the LLM provides too much long information, and the efficiency is not good.

For the social needs of human pilots, the expert panel agreed that the developed V-CoP cannot meet the requirements of being a companion and more efforts are needed to enhance its soft skills.

TABLE III. V-COP PERFORMANCE IN FORMAL EXPERIMENT

| Settings | Accuracy | | |
|---|---|---|---|
| | *IFC[a]* | *GP[b]* | *IC[c]* |
| Image + Pilot's Instruction | 90.5% | 85.5% | 70.5% |

[a.] IFC: Interpreting the flight condition. [b.] GP: Generated procedures; [c.] IC: Index correction

### C. Error Analysis of the V-CoP

To gain further insights into the nature of the errors and enhance the V-CoP's performance in the future, we performed an in-depth analysis of the error types as follows:

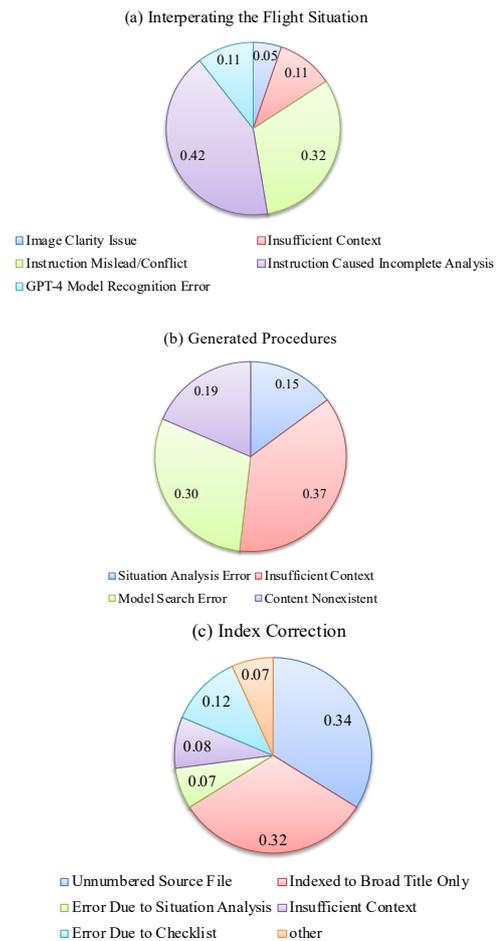

Fig.7. The error analysis of the V-CoP.

For understanding the flight situation, the V-CoP mainly made errors due to the following reasons, including incomplete image analysis, misleading instruction, insufficient context, image clarity, and GPT-4 model recognition error. Specifically, the most significant error type is incomplete image analysis (42.11%), where the V-CoP model did not fully analyze the image content, as shown in Fig.7(a). For the generated procedures, four types of errors were made by the V-CoP, including model search error, insufficient context, nonexistent content, and situation analysis error, as shown in Fig.7(b). Among them, insufficient context, which indicates the provided information was too vague or general for a specific checklist item, made up most errors (37%). For the last aspect, there are many reasons that the V-CoP cannot provide a correct index, such as unnumbered source file, indexed to broad title only, situation analysis error, and insufficient context, as shown in Fig.7(c).

The formal experiment results provide valuable insights into the LLM-based V-CoP's current capabilities and limitations as a pilot assistance system. The performance across different performance evaluators indicates that while the model is proficient in interpreting aviation-related images, certain areas require improvements to enhance accuracy and reliability. The investigation of errors sheds light on the specific challenges faced by the V-CoP:

- Model search errors and content nonexistence imply limitations within the knowledge base and retrieval processes.
- Situation analysis errors and incomplete image analysis suggest issues with image processing and contextual understanding.
- Insufficient context across all categories underscores a need for better integration of the data provided with the model's existing knowledge.

### D. Recommendations for Potential Improvements

During the experiments, we found that a structured and streamlined knowledge base can reduce search errors and content nonexistence issues. By organizing the data in a more accessible and hierarchical manner, the retrieval process can become more efficient and accurate. In this study, the raw and long manuals from Airbus were adopted, which hinders the indexing and retrieval processes. We may potentially employ database management techniques to improve indexing and retrieval processes in the next step.

In addition to the challenges associated with the knowledge base, the V-CoP's performance is significantly influenced by the formulation of instructions, as evidenced by errors arising from misleading directives and inadequate context. Enhanced clarity and precision in instructions could potentially improve situational analysis and indexing. In this research, we utilized basic instructions such as "check the error" and "focus on the yellow part." As a subsequent step, we propose referencing aviation standard phrases to devise a method for generating appropriate prompts for the V-CoP. The development of a standardized protocol for instruction formulation could mitigate ambiguity. This may necessitate the establishment of a guideline set for instruction composition, with an emphasis on clarity and specificity.

Last, providing additional context could mitigate a significant number of errors, particularly those related to insufficient context and incomplete image analysis. The question is how to collect supplementing inputs with background information relevant to the scenario. This might include data about the phase of flight, weather conditions, or even historical data patterns that could enhance the V-CoP's context understanding.

### V. CONCLUSION

This research introduces and delineates the concept of V-CoP, a system that employs a multimodal LLM to interpret and generate human-like text from cockpit images and pilot inputs, thereby offering real-time support during flight operations. To the best of our knowledge, this is the first work to study the virtual co-pilot with pretrained LLMs for aviation. A case study was undertaken to partially explore the feasibility of an LLM-enabled V-CoP in identifying comprehensive, dynamic, and interactive procedures adaptable to various flight scenarios. The case study revealed that GPT-4, when provided with instrument images and pilot instructions, can effectively retrieve quick-access references for flight operations. The findings affirmed that the V-CoP can harness the capabilities of LLM to comprehend dynamic aviation scenarios and pilot instructions.

However, the V-CoP's performance is yet to meet the stringent aviation safety standards. The error analysis indicates the need for improvements in context comprehension, knowledge base refinement, and image analysis capabilities.

REFERENCES

[1] P. L. Myers III and A. W. Starr Jr, "Single pilot operations IN commercial cockpits: background, challenges, and options," *J Intell Robot Syst*, vol. 102, no. 1, p. 19, 2021.
[2] D. Harris, "Single-pilot airline operations: Designing the aircraft may be the easy part," *The Aeronautical Journal*, pp. 1–21, 2023.
[3] A. K. Faulhaber, M. Friedrich, and T. Kapol, "Absence of pilot monitoring affects scanning behavior of pilot flying: implications for the design of single-pilot cockpits," *Hum Factors*, vol. 64, no. 2, pp. 278–290, 2022.
[4] M. L. Cummings, A. Stimpson, and M. Clamann, "Functional requirements for onboard intelligent automation in single pilot operations," in *AIAA Infotech@ Aerospace*, 2016, p. 1652.
[5] E. Kasneci *et al.*, "ChatGPT for good? On opportunities and challenges of large language models for education," *Learn Individ Differ*, vol. 103, p. 102274, 2023.
[6] R. M. Belbin and V. Brown, *Team roles at work*. Routledge, 2022.
[7] T. Inagaki, "Smart collaboration between humans and machines based on mutual understanding," *Annu Rev Control*, vol. 32, no. 2, pp. 253–261, 2008.
[8] R. Mallick, C. Flathmann, C. Lancaster, A. Hauptman, N. McNeese, and G. Freeman, "The pursuit of happiness: the power and influence of AI teammate emotion in human-AI teamwork," *Behaviour & Information Technology*, pp. 1–25, 2023.
[9] X. Cheng, and S. Zhang, Tool, Teammate, Superintelligence: Identification of ChatGPT-Enabled Collaboration Patterns and their Benefits and Risks in Mutual Learning, 2024.
[10] J. V Dinh and E. Salas, "Factors that influence teamwork," *The Wiley Blackwell handbook of the psychology of team working and collaborative processes*, pp. 13–41, 2017.
[11] F. Dehais, J. Behrend, V. Peysakhovich, M. Causse, and C. D. Wickens, "Pilot flying and pilot monitoring's aircraft state awareness during go-around execution in aviation: A behavioral and eye tracking study," *Int. J. Aerosp. Psychol.*, vol. 27, no. 1–2, pp. 15–28, 2017.
[12] F. Li, C.-H. Chen, C.-H. Lee, and L.-P. Khoo, "A user requirement-driven approach incorporating TRIZ and QFD for designing a smart vessel alarm system to reduce alarm fatigue," *The Journal of Navigation*, vol. 73, no. 1, pp. 212–232, 2020.
[13] M. Clamann and D. B. Kaber, "Applicability of usability evaluation techniques to aviation systems," *Int J Aviat Psychol*, vol. 14, no. 4, pp. 395–420, 2004.
[14] A. AIRBUS, "Aircraft characteristics airport and maintenance planning." AIRBUS SAS, 2017.
[15] Airbus, "Flight Crew Operating Manual A318/A319/A320/A321 FMGS Pilot's Guide Vol 4," https://www.avialogs.com/aircraft-

a/airbus/item/1004-flight-crew-operating-manual-a318-a319-a320-a321-fmgs-pilot-s-guide-vol-4.